\documentclass{elsart3p}

\usepackage{graphicx}

\usepackage{amssymb}

\begin{document}

\begin{frontmatter}



\title{Theoretical study of NMR relaxation due to rattling phonons}


\author{Thomas Dahm \corauthref{cor1}\thanksref{label2}}
\ead{dahm@uni-tuebingen.de}
\author{and Kazuo Ueda}
\address{Institute for Solid State Physics, University of Tokyo, 
Kashiwanoha, Kashiwa, Chiba 277-8581, Japan}
\corauth[cor1]{Corresponding author.}
\thanks[label2]{On leave of absence from Institut f\"ur Theoretische Physik, 
         Universit\"at T\"ubingen, Germany}

\begin{abstract}
We calculate the NMR relaxation rate due to quadrupolar coupling of
the nucleus to a local, strongly anharmonic phonon mode. As a model
potential for a ``rattling'' motion we consider a square-well
potential. We calculate the free phonon Green's function
analytically and derive the low and high temperature limits
of the NMR relaxation rate. It is shown that the temperature
dependence of the NMR relaxation rate possesses a peak in
contrast to harmonic phonons but in qualitative agreement
with a recent NMR study on KOs$_2$O$_6$. We discuss the
influence of phonon renormalization due to electron-phonon
interaction.
\end{abstract}

\end{frontmatter}


The new pyrochlore oxide superconductor KOs$_2$O$_6$ shows unusual 
behavior both in the normal as well as in the superconducting state. 
Density functional calculations have shown that the vibrations of the 
potassium ion are highly anharmonic, allowing large excursions from 
its equilibrium position within the surrounding Os-O cage \cite{Kunes1},
consistent with recent X-ray studies \cite{Yamaura}.
Such a situation, in which a small ion can move anharmonically within 
an oversized cage of surrounding atoms has been called ``rattling''
and also exists in other compounds \cite{Keppens,Paschen,Nakai}.
The anharmonic dynamics of the potassium ion in KOs$_2$O$_6$
is currently being 
discussed as a possible origin for the anomalies seen in various 
experimental quantities, like e.g. specific heat \cite{Hiroirev}, 
resistivity \cite{Hiroi}, thermal
conductivity \cite{Kasahara}, and NMR relaxation \cite{Yoshida}. 

Here we focus on recent NMR studies of KOs$_2$O$_6$. 
In particular, the nuclear spin-lattice relaxation 
rate $1/T_1 T$ at the potassium site can directly probe the unusual 
potassium dynamics through quadrupolar coupling of the K nucleus 
to the electric field gradient, as has been shown recently \cite{Yoshida}.
The temperature dependence of the NMR relaxation 
rate $1/T_1 T$ was found to be highly anomalous, increasing at low
temperatures, reaching a peak value around $\sim 14$~K and decreasing
again at higher temperatures. Such a behavior is inconsistent with
the $1/T_1 \sim T^2$ behavior expected from harmonic phonons at
high temperatures due to the two-phonon Raman process \cite{Abragam}.

In a previous work we succeeded in describing this anomalous temperature
dependence considering a strong anharmonicity of the vibration of
the potassium ion \cite{DahmUeda}. We added a fourth order term to
the harmonic potential and studied its influence within a self-consistent
quasi-harmonic approximation. Within this model the effective phonon
frequency becomes a strongly increasing function of temperature,
modifying the higher temperature behavior of both NMR relaxation
rate and resistivity \cite{DahmUeda}, consistent with the
experimental observations.

In the present work we are studying a square-well potential as
a model for the potassium dynamics. This potential may
seem less realistic and it also does not allow to ``tune'' the
amount of anharmonicity. However, it can be regarded as the most
extreme case of ``rattling'' and, more importantly, the
square-well potential allows us to calculate the phonon
spectral function analytically, without having to resort
to the quasi-harmonic approximation. We wish to demonstrate
that an anomalous temperature dependence of the NMR relaxation
rate like the one found in Ref.~\cite{DahmUeda} results,
reinforcing our previous conclusion.

As the simplest possible model we consider the one-dimensional
square-well potential with infinitely high walls.
As is well known, the energy levels in this case are given by
\begin{equation}
E_n = \frac{\hbar^2 \pi^2}{2 m L^2} n^2 \qquad \mbox{with} \qquad
n = 1,2,\cdots
\label{energylevels}
\end{equation}
Here, $L$ is the size of the well and $m$ the mass of the atom.
It is an easy exercise to analytically calculate all the matrix
elements of the positional operator $x$, giving
\begin{equation}
\langle n | x | m \rangle = \frac{2L}{\pi^2} \left[ \frac{1}{\left( m + n \right)^2} 
- \frac{1}{\left( m - n \right)^2} \right] ,
\end{equation}
if $m+n$ odd and 0 otherwise.
The free, non-interacting phonon Green's function $D_0(\omega)$
can be calculated quite generally from the expression \cite{Rickayzen}
\begin{equation}
D_0(\omega) = \frac{1}{Z} \sum_{m,n}
\frac{\left( e^{-\beta E_m} -  e^{-\beta E_n} \right) 
\left| \langle n | x | m \rangle \right|^2}{
\left( \omega - E_n + E_m \right) \left| \langle 1 | x | 2 \rangle \right|^2}
\label{freephonon}
\end{equation}
Here, $Z=\sum_{n} e^{-\beta E_n}$ is the partition function and 
$\beta=1/T$ the inverse temperature and we have normalized $D_0(\omega)$
to the lowest energy excitation $\langle 1 | x | 2 \rangle $.

In the superconducting state the NMR relaxation rate at the potassium
site exhibits a sudden decrease. As has been pointed out in 
Ref.~\cite{Yoshida} this suggest that the rattling phonon must
be strongly coupled to the conduction electrons. Following
Ref.~\cite{DahmUeda} we therefore introduce a finite phonon
self energy $\Pi(\omega)$ due to coupling to the conduction
electrons. Then, the interacting phonon Green's function $D(\omega)$
can be calculated from Dyson's equation
\begin{equation}
D(\omega) = \frac{1}{D_0^{-1} \left( \omega \right) - 
\Pi \left( \omega \right)} 
\end{equation}
The phonon spectral function $A(\omega)$ is obtained from
\begin{equation}
A(\omega) = - \frac{1}{\pi} \, \mbox{Im} \, D(\omega + i 0^+) 
\label{specfunc}
\end{equation}
Because the electronic energy scale is much larger than the
phononic energy scale, we can make a low energy expansion
of the phonon self energy and write
\begin{equation}
\Pi (\omega) \approx - \Pi_1 - i \alpha \omega
\label{selfenergy}
\end{equation}
where $\Pi_1$ and $\alpha$ are positive real constants describing the
interaction of the phonon with the conduction electrons. Here,
$\Pi_1$ is the renormalization of the phonon frequency and
$\alpha$ determines the phonon damping rate.

\begin{figure}[t] 
  \begin{center}
    \includegraphics[width=0.75\columnwidth,angle=270]{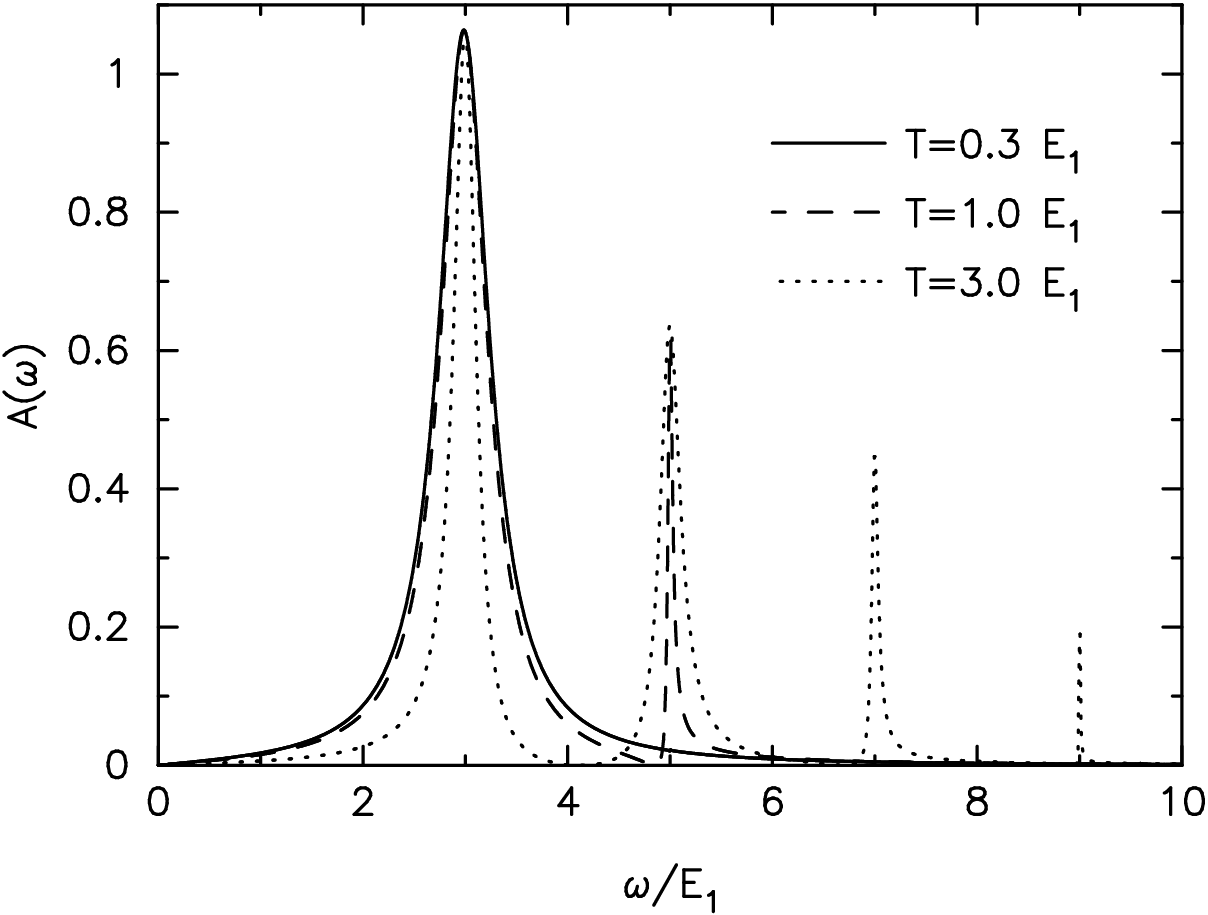}
    \caption{ Phonon spectral function $A(\omega)$ for the square well
potential at three different temperatures $T=0.3 E_1$ (solid line),
$T=1.0 E_1$ (dashed line), and $T=3.0 E_1$ (dotted line).
With increasing temperature, spectral weight is shifted gradually
away from the main peak at $3E_1$ to the higher energy peaks.
    \label{fig1} }
  \end{center}
\end{figure} 

As can be seen from Eq.~(\ref{freephonon}) the non-interacting
phonon Green's function consists of a sum of poles at the
transition energies $E_n-E_m$. For a harmonic oscillator these
poles all fall on top of each other, resulting in a single,
temperature independent pole. However, for the present square-well
potential the allowed transition energies are different and the
corresponding spectral function consists of a series of peaks
at the positions $\delta_n=\frac{\hbar^2 \pi^2}{2 m L^2} (2n+1)$.
The relative weight of these peaks is temperature dependent.
With increasing temperature spectral weight is gradually transfered
to higher energies, because higher energy transfers become
possible. In Fig.~\ref{fig1} we show the energy dependence of
the spectral function $A(\omega)$ calculated from 
Eqs.~(\ref{energylevels})-(\ref{specfunc}) at
three different temperatures $T=0.3 E_1$ (solid line),
$T=1.0 E_1$ (dashed line), and $T=3.0 E_1$ (dotted line).
Here, $E_1=\frac{\hbar^2 \pi^2}{2 m L^2}$ is the ground state
energy of the square well potential. For illustration, no phonon
frequency renormalization $\Pi_1=0$ and a small phonon
damping of $\alpha=0.1$ has been chosen. The influence of
these two parameters on the NMR relaxation rate will be
discussed below. Fig.~\ref{fig1} illustrates
how the spectral weight is shifted away from the lowest energy
peak at $3E_1$ to the higher energy peaks, when temperature is
increasing.

Within the quasi-harmonic approximation of Ref.~\cite{DahmUeda}
this gradual transfer of spectral weight to higher energies
appears as a temperature increase of the effective phonon
frequency. We wish to emphasize that the peak positions do
not change as a function of temperature. Only their intensities
do. The effective phonon frequency of Ref.~\cite{DahmUeda}
therefore corresponds to an effective average frequency and
should not be associated with an energy level distance. For example,
a neutron scattering experiment is expected to observe a
temperature dependent intensity of a series of phonon peaks and will
not be able to directly observe the effective average frequency.

\begin{figure}[t] 
  \begin{center}
    \includegraphics[width=0.75\columnwidth,angle=270]{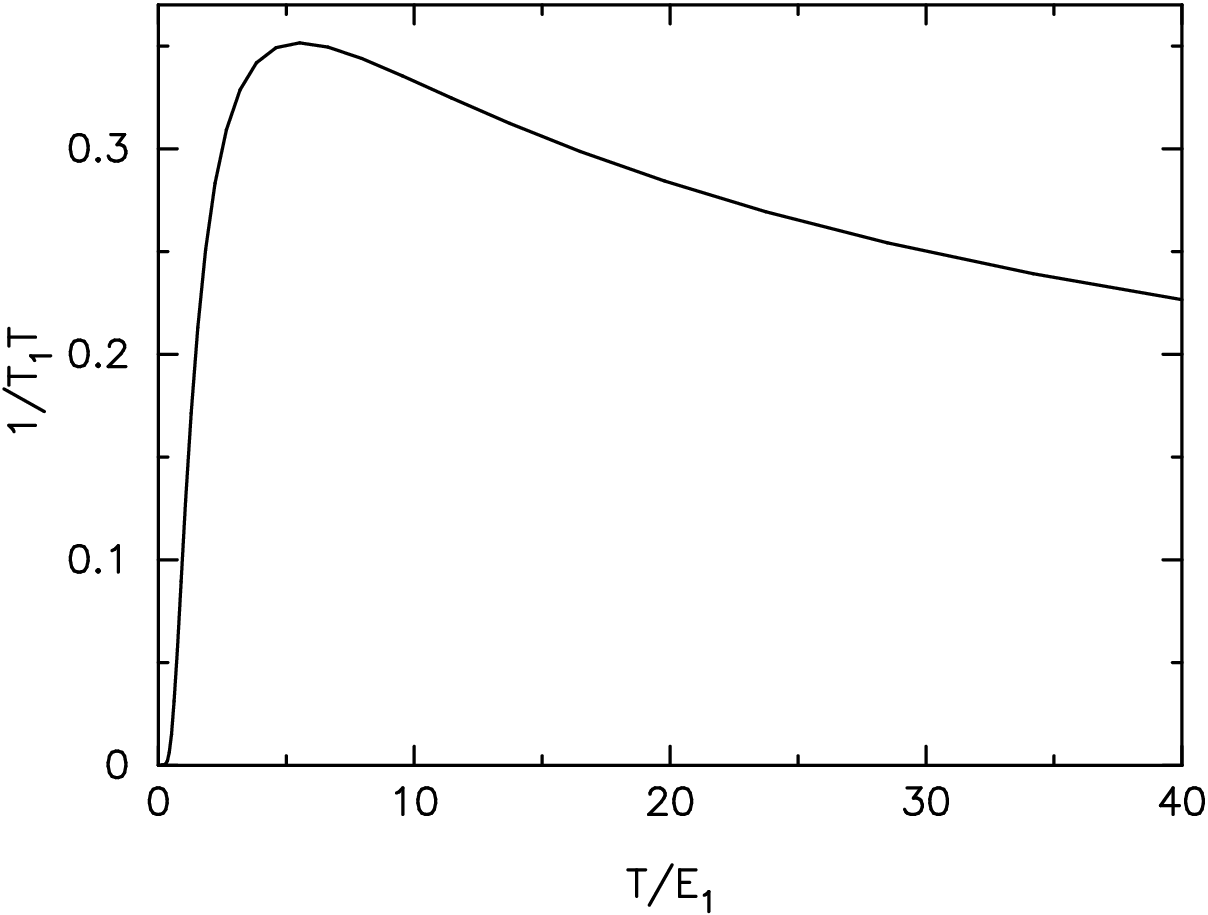}
    \caption{ Temperature dependence of the NMR relaxation rate $1/T_1T$
from the two-phonon Raman process for the square-well potential.
    \label{fig2} }
  \end{center}
\end{figure}

Once the phonon spectral function is known, the NMR relaxation
rate $1/T_1$ from quadrupolar interaction of the potassium
nucleus to the local electric field gradient can be calculated.
In Ref.~\cite{DahmUeda} we have shown that the two-phonon Raman
process is dominating over the direct process, because of the
small number of phonon states available at the Larmor frequency
\cite{Abragam}. Therefore, here we will focus on the two-phonon 
Raman process. In this case $1/T_1$ is given by
\begin{equation}
\frac{1}{T_1} \propto
\int_{-\infty}^\infty d\omega \; A^2 (\omega)
\left[ n( \omega ) + 1 \right] n (\omega )
\label{Ramanprocess}
\end{equation}
neglecting higher order vertex corrections due to electron-phonon
coupling. Here, $n (\omega )$ is the Bose distribution function.
In Fig.~\ref{fig2} we show a numerical evaluation of Eq.~(\ref{Ramanprocess})
using the spectral function from Eq.~(\ref{specfunc}). Again,
$\Pi_1=0$ and $\alpha=0.1$ has been chosen. It is immediately
apparent that the NMR relaxation rate $1/T_1T$ possesses a peak
as a function of temperature near $T=5 E_1$, in contrast to harmonic phonons.
The behavior is qualitatively similar to the experimental data
from Ref.~\cite{Yoshida}. At low temperatures $1/T_1T$ varies
like $T^2$ due to the linear behavior of the spectral function
at low energies seen in  Fig.~\ref{fig1}. At high temperatures
we can show that $1/T_1T$ varies like $1/\sqrt{T}$ (see below). 
This is in contrast to
harmonic phonons, where $1/T_1T \sim T$ at high temperatures,
but it is also different from the fourth order potential result
where $1/T_1T \sim $~const was found \cite{DahmUeda}.
Apparently the high temperature behavior depends somewhat on
the type of potential chosen.

In order to derive the high temperature limiting behavior we assume
that the damping parameter $\alpha$ is small enough such that the
poles appearing in Eq.~(\ref{freephonon}) remain well separated.
Generally we can write $D_0(\omega)$ as a series of poles:
\begin{equation}
D_0(\omega) = \sum_n \frac{a_n \left( T \right)}{\omega-\delta_n}
\end{equation}
where $a_n \left( T \right)$ are the temperature dependent
weights of the poles. Because the matrix elements
$\left| \langle n | x | m \rangle \right|^2$ quickly decrease
with increasing distance $n-m$, to a good approximation we can
only keep the nearest neighbor terms $m=n \pm 1$. Then we
have
\begin{equation}
a_n \left( T \right) = 
\frac{ e^{-\beta E_{n \pm 1}} -  e^{-\beta E_{n}} }{Z\left( T \right)}
\label{an}
\end{equation}
The partition function $Z(T)=\sum_{n} e^{-\beta E_1 n^2}$ varies
like $\sqrt{T}$ at high temperatures because the number of
terms contributing significantly to the sum increases as $\sqrt{T}$.
If the poles are well separated, the full phonon Green's function
becomes
\begin{equation}
D(\omega) \approx \sum_n \frac{a_n \left( T \right)}{\omega-\delta_n
- a_n \left( T \right) \Pi \left( \omega \right)}
\end{equation}
and the spectral function can be written in the form
\begin{equation}
A(\omega) \approx \frac{1}{\pi} \sum_n 
\frac{a_n \left( T \right) \Gamma_{n,eff}}{\left[ \omega-\delta_n
+ a_n \left( T \right) \Pi_1 \right]^2 + \Gamma_{n,eff}^2}
\label{specfuncapprox}
\end{equation}
with $\Gamma_{n,eff}=\alpha \, a_n \left( T \right)
\left( \delta_n - a_n \left( T \right) \Pi_1 \right)$.
Using Eq.~(\ref{specfuncapprox}) and evaluating each pole
separately, we find from Eq.~(\ref{Ramanprocess})
\begin{equation}
\frac{1}{T_1}
\approx \frac{1}{2\pi} \sum_n 
\frac{a_n \left( T \right) }{\alpha \, \Omega_n}
\left[ n( \Omega_n ) + 1 \right] n (\Omega_n )
\end{equation}
where $\Omega_n=\delta_n - a_n \left( T \right) \Pi_1$ are 
the renormalized phonon frequencies. Apparently, $1/T_1$ scales
like $1/\alpha$ for all temperatures and small values of $\alpha$.
This means that the parameter $\alpha$ has no large influence on the
temperature dependence of the NMR relaxation rate, but only
on its absolute values.

In the high temperature limit we find from Eq.~(\ref{an})
\begin{equation}
a_n \left( T \right) \approx
\frac{ \beta \left( E_{n} - E_{n \pm 1} \right) }{Z\left( T \right)}
\sim T^{-3/2}
\end{equation}
Using $n (\Omega_n ) \sim T/\Omega_n $ we finally arrive at
\begin{equation}
\frac{1}{T_1T} \propto
\frac{1}{\alpha \sqrt{T}} \, .
\end{equation}

\begin{figure}[t] 
  \begin{center}
    \includegraphics[width=0.75\columnwidth,angle=270]{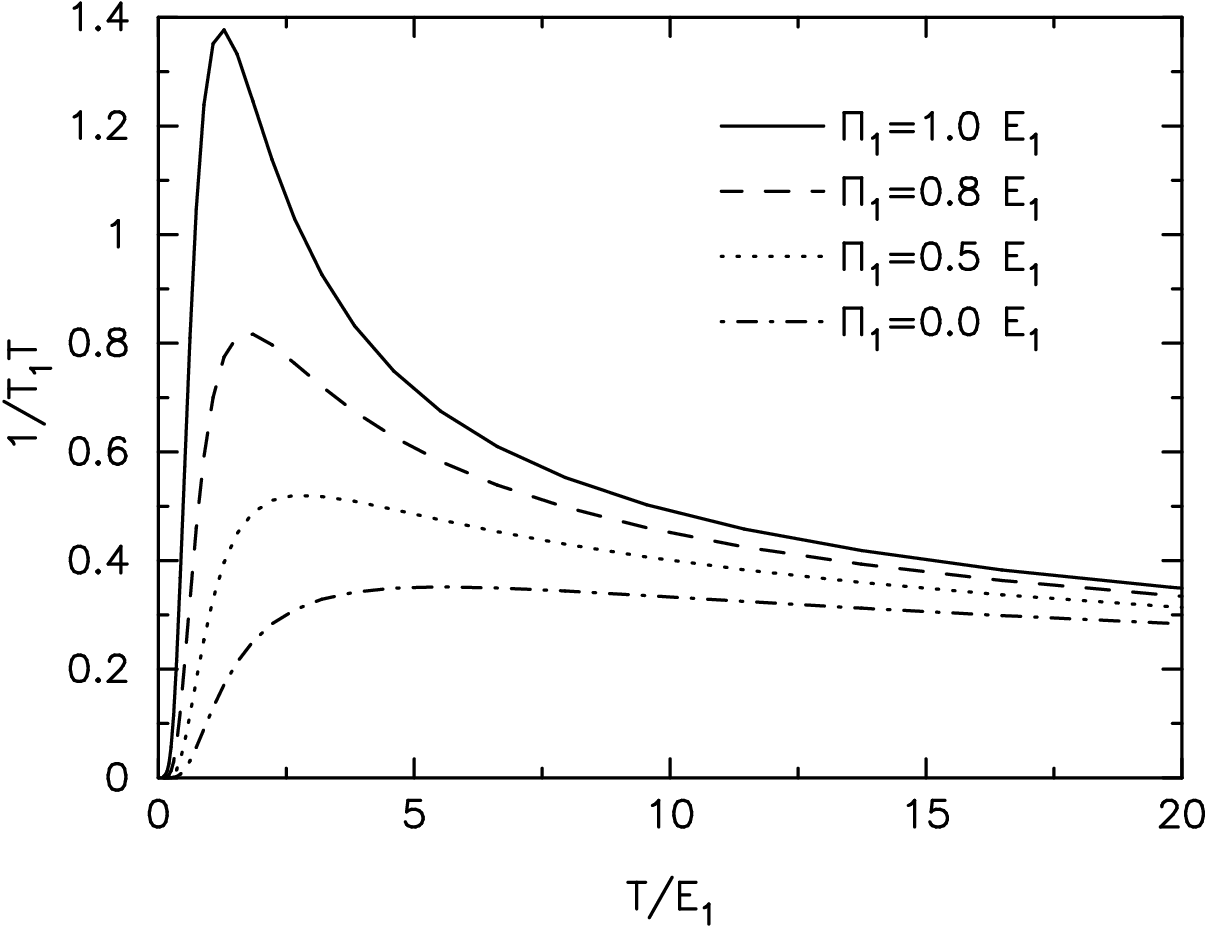}
    \caption{ Temperature dependence of the NMR relaxation rate $1/T_1T$
from the two-phonon Raman process for different values of the real part 
of the phonon self-energy: $\Pi_1=E_1$ (solid line), 
$\Pi_1=0.8 E_1$ (dashed line), $\Pi_1=0.5 E_1$ (dotted line), and 
$\Pi_1=0$ (dashed-dotted line).
    \label{fig3} }
  \end{center}
\end{figure}

We can estimate an absolute value of the expected peak position 
of the NMR relaxation rate by choosing
the potassium mass for $m$ in Eq.~(\ref{energylevels}) and a size of the
well of $L \approx 1$~\AA, as suggested by both bandstructure calculations
\cite{Kunes1} and X-ray studies \cite{Yamaura}. Then we find 
$E_1 \approx 6.1$~K, which would suggest the NMR peak position to
appear around $5E_1 \approx 30$~K. However, the peak in the experimental data 
appears near 14~K and also the peak is somewhat more sharp than the
one in Fig.~\ref{fig2}. A possible explanation for these
differences could be the influence of a sizeable phonon 
renormalization due to electron-phonon interaction, as given
by the parameter $\Pi_1$ in Eq.~(\ref{selfenergy}). To
illustrate the influence of the parameter $\Pi_1$, in Fig.~\ref{fig3}
we show the temperature dependence of $1/T_1T$ for four different
values of $\Pi_1$, keeping $\alpha=0.1$ fixed. 
With increasing $\Pi_1$ the peak becomes more
sharp and the peak position moves to lower temperatures easily
softening the peak position by more than a factor of 2.
Thus, a better agreement with the NMR data on KOs$_2$O$_6$
could be obtained for a value of $\Pi_1 \approx 0.5 E_1$.

To summarize we have calculated the NMR relaxation rate due to
quadrupolar coupling to a local, strongly
anharmonic phonon mode considering the two-phonon Raman process.
We have calculated the free phonon Green's function analytically
for a square-well potential. Including phonon renormalization
due to electron-phonon interaction we were able to show that
the temperature dependence of the NMR relaxation rate $1/T_1T$
possesses a peak as a function of temperature, in stark contrast
to harmonic phonons. The position and height of this peak depends
on the strength of the phonon renormalization. We also derived the
high and low temperature limiting behaviors. These results
reinforce our previous approximate treatment of a fourth order 
potential.

We would like to thank Z.~Hiroi, K.~Ishida, Y.~Matsuda, T.~Shibauchi, and
M.~Takigawa for valuable discussions.


\begin{thebibliography}{00}




\bibitem{Kunes1} J.~Kunes, T.~Jeong, and W.~E.~Pickett, Phys. Rev.
B {\bf 70}, 174510 (2004).

\bibitem{Yamaura} J.~Yamaura, S.~Yonezawa, Y.~Muraoka, and Z.~Hiroi,
J. Solid State Chem. {\bf 179}, 336 (2006).

\bibitem{Keppens} V.~Keppens,~D.~Mandrus,~B.~C.~Sales, B.~C.~Chakoumakos, 
P.~Dai, R.~Coldea, M.~B.~Maple, D.~A.~Gajewski, E.~J.~Freeman, and 
S.~Bennington,
Nature (London) {\bf 395}, 876 (1998).

\bibitem{Paschen} S.~Paschen, W.~Carrillo-Cabrera, A.~Bentien, 
V.~H.~Tran, M.~Baenitz, Yu.~Grin, and F.~Steglich, Phys. Rev.
B {\bf 64}, 214404 (2001).

\bibitem{Nakai} Y.~Nakai, K.~Ishida, K.~Magishi, H.~Sugawara, D.~Kikuchi,
and H.~Sato, J. Magn. Magn. Mater. {\bf 310}, 255 (2007).

\bibitem{Hiroirev} Z.~Hiroi, S.~Yonezawa, Y.~Nagao and J.~Yamaura, 
Phys. Rev. B {\bf 76}, 014523 (2007).

\bibitem{Hiroi} Z.~Hiroi, S.~Yonezawa, J.~Yamaura, T.~Muramatsu,
and Y.~Muraoka, J. Phys. Soc. Jpn. {\bf 74}, 1682 (2005).

\bibitem{Kasahara} Y.~Kasahara, Y.~Shimono, T.~Shibauchi, Y.~Matsuda,
S.~Yonezawa, Y.~Muraoka, and Z.~Hiroi,
Phys. Rev. Lett. {\bf 96}, 247004 (2006).

\bibitem{Yoshida} M.~Yoshida, K.~Arai, R.~Kaido, M.~Takigawa, 
S.~Yonezawa, Y.~Muraoka, and Z.~Hiroi,
Phys. Rev. Lett. {\bf 98}, 197002 (2007).

\bibitem{Abragam} A.~Abragam, {\it The principles of nuclear relaxation} 
(Oxford University Press, Oxford, 1961).

\bibitem{DahmUeda} T.~Dahm and K.~Ueda, 
Phys. Rev. Lett. {\bf 99}, 187003 (2007).

\bibitem{Rickayzen} G.~Rickayzen, {\it Green's Functions and Condensed 
Matter}, (Academic Press, London 1980).

\end{thebibliography}
\end{document}